\documentclass[a4paper]{jpconf}
\usepackage{graphicx}
\begin{document}
\title{Quasiparticles of string solutions in the spin-1/2 antiferromagnetic Heisenberg chain in a magnetic field}

\author{Masanori Kohno}

\address{International Center for Materials Nanoarchitectonics (MANA), 
National Institute for Materials Science (NIMS), Tsukuba 305-0044, Japan}

\ead{KOHNO.Masanori@nims.go.jp}

\begin{abstract}
Spectral properties of the spin-1/2 antiferromagnetic Heisenberg chain in a magnetic field 
are investigated by using exact Bethe-ansatz solutions. We argue that not only quasiparticles called 
psinon and antipsinon but also a quasiparticle representing a 2-string in the Bethe ansatz plays 
an important role for dynamical properties in a magnetic field. Combined with psinon and antipsinon, 
the quasiparticle for a 2-string forms a continuum in the high-energy regime for transverse dynamical structure 
factor $S^{+-}(k,\omega)$. In the zero-field limit, the continuum is located on the mode of the lowest excited states 
in zero field called the des Cloizeaux-Pearson mode. In a magnetic field, the continuum separates 
from other low-energy continua, and reduces to the mode of bound states of overturned spins 
from the fully polarized state near the saturation field. We confirm the relevance through comparisons 
with available experimental results on the quasi-one-dimensional antiferromagnet CuCl$_2\cdot$2N(C$_5$D$_5$). 
\end{abstract}

\section{Introduction}
It is known that dynamical properties of usual antiferromagnets with antiferromagnetic orders are well explained in terms of magnons 
in spin-wave theory \cite{SWAF}. 
On the other hand, dynamical properties in the one-dimensional (1D) spin-1/2 antiferromagnetic Heisenberg chain 
are known to be characterized by quasiparticles (QPs) called spinons \cite{FT,HSspinon} in the absence of a magnetic field. 
These QPs are different in various aspects, such as quantum numbers, lineshapes in dynamical structure factors (DSFs), 
and definitions. Namely, the former is defined as a quantum fluctuation around classical spin configurations, 
and carries an integer magnetization, showing a $\delta$-functional lineshape in DSFs. 
In contrast, the latter is defined in the Bethe-ansatz formulation \cite{Bethe}, 
and carries a fractional quantum number, showing power-law behaviors in DSFs. 
\par
In this paper, we consider QPs of the 1D Heisenberg chain in the presence of a magnetic field. 
Recently, Karbach and his coworkers have introduced QPs called psinon and anti-psinon 
as a hole and a particle in the distributions of Bethe quantum numbers \cite{Bethe} in a magnetic field \cite{Karbach_psinon,Karbach_Szz}. 
These QPs well explain dynamical properties of the Heisenberg chain for DSFs 
$S^{-+}(k,\omega)$ \cite{Karbach_psinon} and $S^{zz}(k,\omega)$\cite{Karbach_psinon,Karbach_Szz}. 
However, the behaviors in $S^{+-}(k,\omega)$ have not been fully understood in terms of these QPs. 
For example, numerical simulations have suggested that there is a mode at high energies \cite{Muller,Lefmann} 
which does not follow the behaviors predicted based on the Bethe formalism \cite{Muller}. 
Also, experimentally observed high-energy peaks \cite{CPC,CuPzN} have not been quantitatively explained by psinons and anti-psinons. 
This implies that we also need to consider other QPs to explain whole dynamical properties of 1D chains especially for $S^{+-}(k,\omega)$. 

\section{Model and method}
We consider the spin-1/2 antiferromagnetic Heisenberg chain in a magnetic field under the periodic boundary condition 
${\mbox {\boldmath $S$}}_{L+1}={\mbox {\boldmath $S$}}_1$. The Hamiltonian is defined by 
\begin{equation}
{\cal H}=J\sum_{j=1}^{L}{\mbox {\boldmath $S$}}_j\cdot{\mbox {\boldmath $S$}}_{j+1}-HS^z, 
\label{eq:1}
\end{equation}
where ${\mbox {\boldmath $S$}}_j$ is the spin-1/2 operator at site $j$, and $S^z\equiv\sum_jS^z_j$. 
The coupling constant $J$ is positive, reflecting the antiferromagnetic interaction. 
The length of the chain is denoted by $L$. 
The magnetic field $H$ in the thermodynamic limit has been obtained in Ref. \cite{Griffiths}. 
\par
We investigate dynamical properties using exact Bethe-ansatz solutions. 
In the Bethe ansatz \cite{Bethe}, the wavefunction of an eigenstate is expressed in a plane-wave form as
\begin{equation}
\Phi(x_1,\cdots,x_M)=\sum_{P}\exp\left[\i\sum_{j=1}^M{\tilde k}_{Pj}x_j
+\i\sum_{\tiny \begin{array}{c}j<l\\Pj>Pl\end{array}}^M\phi_{PjPl}\right], 
\label{eq:WF}
\end{equation}
where $x_j$ is the site of the $j$-th down spin for $x_1<x_2<\cdots<x_M$, 
and $M$ and $P$ denote the number of down spins and permutations of $1,\cdots,M$, respectively. 
Here, $\phi_{jl}$ and ${\tilde k}_j$ are given as 
\begin{equation}
2\cot\frac{\phi_{jl}}{2}=\cot\frac{{\tilde k}_j}{2}-\cot\frac{{\tilde k}_l}{2} \quad \mbox{and} \quad {\tilde k}_j=\pi-2\arctan \Lambda_j, 
\label{eq:quasimomenta}
\end{equation}
where rapidity $\Lambda_j$ is obtained from the Bethe equation \cite{Bethe}:
\begin{equation}
L\arctan \Lambda_j=\pi I_j+\sum_{l=1}^M\arctan\frac{\Lambda_j-\Lambda_l}{2}.
\label{eq:quasimomenta}
\end{equation}
Here, $\{I_j\}$ is a set of integers or half-odd integers, called Bethe quantum numbers. 
The above formulation means that the wavefunction of an eigenstate is obtained, once $\{I_j\}$ is given. 
The distribution of $\{I_j\}$ is somewhat analogous to the momentum distribution of the spinless fermion model in a chain. 
Namely, in the ground state in a magnetic field, $I_j$'s are densely distributed around zero like a Fermi sea. 
Excitations from the dense distribution can be obtained by creating holes inside it or particles outside it. 
The hole and the particle are called psinon and anti-psinon, and denoted by $\psi$ and $\psi^*$, respectively \cite{Karbach_psinon}. 
\par
Here, we consider solutions with two complex rapidities: $\Lambda_j$ for $j=1,\cdots, M-2$ are real, 
$\Lambda_{M-1}={\bar \Lambda}+\i+c_1$, and $\Lambda_M={\bar \Lambda}-\i+c_2$ with real ${\bar \Lambda}$, 
where $c_1$ and $c_2$ are numbers which become exponentially small in the $L\rightarrow\infty$ limit. 
The pair of the complex rapidities is called a string of length two or a 2-string. For these solutions, we introduce two sets of $\{I_j\}$ \cite{Takahashi}. 
One is for the real rapidities ($\Lambda_j$ for $j=1,\cdots, M-2$), and the other is for the string (${\bar \Lambda}$). 
We denote them by $\{I_j^{\rm r}\}$ and  $\{I_j^{\rm i}\}$, respectively. Solutions with a string of length three or a 3-string are similarly defined. 
For these complex-rapidity solutions, the particle in the distribution of $\{I_j^{\rm i}\}$ can be regarded as the QP for the 2- or 3-string. 
In this paper, we denote them as $\sigma_2$ and $\sigma_3$. 
\par
We consider behaviors of DSFs defined by 
\begin{equation}
S^{{\bar\alpha}\alpha}(k,\omega)=\sum_i |\langle k,\epsilon_i|S^{\alpha}_k|\mbox{G.S.}\rangle|^2\delta(\omega-\epsilon_i) 
\label{eq:Sqw}
\end{equation}
for $\alpha$=$-$, $z$ and $+$, where $|\mbox{G.S.}\rangle$ and $|k,\epsilon_i\rangle$ denote 
the ground state and an excited state with excitation energy $\epsilon_i$ and momentum $k$ in a magnetic field. 
We also define $S^{\rm av}(k,\omega)$$\equiv$$[S^{-+}(k,\omega)$+$S^{+-}(k,\omega)$+$4S^{zz}(k,\omega)]/6$. 
We calculated DSFs using the Bethe-ansatz solutions, following Refs. \cite{Kitanine, Biegel, Caux}. 
We used real-rapidity solutions of up to 2$\psi$2$\psi^*$ excitations. 
For $S^{zz}(k,\omega)$ and $S^{+-}(k,\omega)$, we took into account 2-string solutions of $O(L^3)$ states as well. 
For $S^{+-}(k,\omega)$, contributions of 3-string solutions of $O(L^3)$ states were also included. 

\section{Results}
\begin{figure}[t]
\begin{center}
\includegraphics[width=16cm]{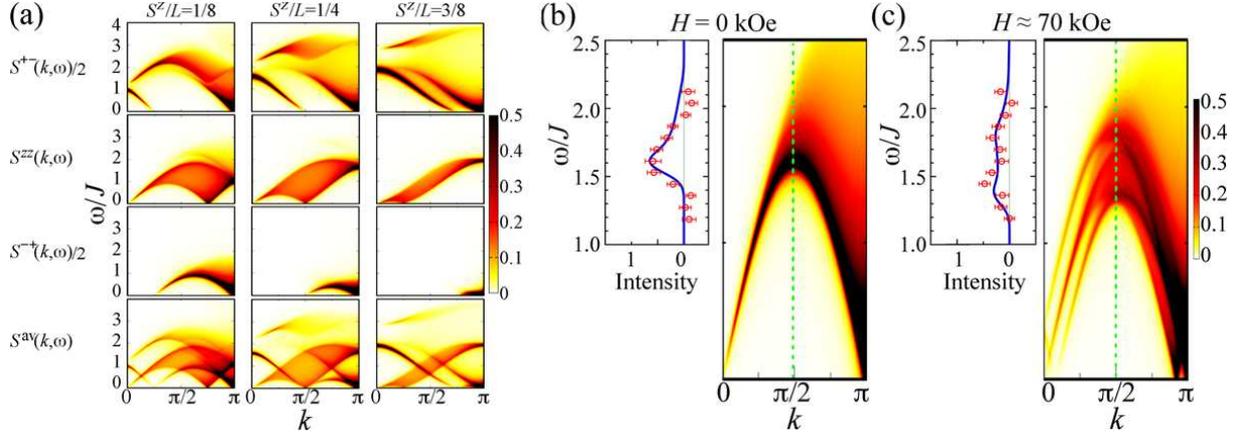}
\end{center}
\caption{(a) Intensity plot of $S^{+-}(k,\omega)$/2, $S^{zz}(k,\omega)$, $S^{-+}(k,\omega)$/2, 
and $S^{\rm av}(k,\omega)$ (from above) at $S^z/L$=1/8, 1/4 and 3/8 (from the left) for $L$=320, 
broadened in a Lorentzian form with full width at half maximum (FWHM) 0.08$J$. 
(b, c) Comparisons with experimental results on CuCl$_2\cdot$2N(C$_5$D$_5$) \cite{CPC} at $k$=$\pi$/2. 
(b) $H$=0 kOe. (c) $H$$\simeq$70 kOe. 
(Left panels in (b,c)) Blue lines are the present results of $S^{\rm av}(k,\omega)$. 
Symbols are the experimental results in Ref. \cite{CPC}. 
The experimental data are rescaled after subtracting the background using $g$-factor $g$=2.08 and $J$=27.32 K \cite{CPCprm}, 
and the present results are broadened in a gaussian form to fit the data in zero field in (b). 
Using the same rescaling and broadening parameters, the results in a magnetic field are compared in (c). 
(Right panels in (b,c))  $S^{\rm av}(k,\omega)$ in $L$=320 broadened as in (a). 
Green dotted lines are the scan paths for the panels on the left. Intensities are shown in units of $1/J$.}
\label{fig:Skw}
\end{figure}

Figure~\ref{fig:Skw} (a) shows the results of DSFs in a magnetic field. We can identify each continuum in terms of the QPs 
as shown in Table \ref{tbl:qp}. 
It should be noted that 2-string solutions also carry large spectral weights and form the high-energy continuum in $S^{+-}(k,\omega)$ \cite{Kohno}. 
In the zero-field limit, this continuum asymptotically reduces to the dominant mode 
in zero field called the des Cloizeaux-Pearson mode \cite{dCP}, because the $I_1^{\rm i}$ for the 2-string is locked at zero 
and the psinon in $\{I_j^{\rm r}\}$ behaves like a spinon. 
On the other hand, near the saturation field, the real rapidities vanish, and the mode of the 2-string reduces to 
that of bound states of two overturned spins from the fully polarized state known for ferromagnetic chains. 
\par
We compare the present results with the experimental results on CuCl$_2\cdot$2N(C$_5$D$_5$) (CPC) in Ref. \cite{CPC} 
as shown in Fig.~\ref{fig:Skw} (b) and (c). The peak width in the present results and the height of the experimental results are adjusted to fit the data in zero field 
as in the left panel of Fig. \ref{fig:Skw} (b). 
Using the same broadening and normalization parameters, we compared the present results with the experimental results at $H$$\approx$70 kOe 
as shown in the left panel of Fig. \ref{fig:Skw} (c). 
This figure shows that the low- and high-energy peaks observed at $H$$\approx$70 kOe are mainly due to the 2$\psi$ continuum in $S^{-+}(k,\omega)$ 
and the $\sigma_2\psi\psi^*$ continuum of 2-string solutions in $S^{+-}(k,\omega)$, respectively. 
\begin{center}
\begin{table}[t]
\caption{Quasiparticles for dynamically dominant excitations in a magnetic field.}
\label{tbl:qp}
\centering
\begin{tabular}{clll}
\br
DSFs&Regions of continua&Quasiparticles&References\\
\br
$S^{+-}(k,\omega)$&
$\begin{array}{l}
k\simeq0, \mbox{low } \omega\\
k\simeq\pi, \mbox{low } \omega\\
k\simeq\pi/4, \mbox{high } \omega\\
k\simeq\pi, \mbox{high } \omega^{4)}\\
\end{array}$
&
$\begin{array}{l}
2\mbox{ }\psi^*\mbox{'s}^{1)}\\
2\mbox{ }\psi^*\mbox{'s}^{2)}\\
\psi, \psi^{*3)} \mbox{ and }\sigma_2\\
2\mbox{ }\psi\mbox{'s} \mbox{ and }\sigma_3
\end{array}$
&
$\begin{array}{l}
\cite{Karbach_psinon,Muller,Biegel,Kohno}\\
\cite{Kohno}\\
\cite{Kohno}\\
\cite{Kohno}
\end{array}$\\
\mr
$S^{zz}(k,\omega)$&
$\begin{array}{l}
k\simeq\pi/2\\
k\simeq\pi, \mbox{high } \omega\\
\end{array}$
&
$\begin{array}{l}
\psi$ and $\psi^*\\
2\mbox{ }\psi\mbox{'s} \mbox{ and }\sigma_2\\
\end{array}$
&
$\begin{array}{l}
\cite{Karbach_psinon,Karbach_Szz,Muller,Biegel,Kohno}\\
\cite{Kohno}
\end{array}$\\
\mr
$S^{-+}(k,\omega)$&
$\begin{array}{l}
k\simeq\pi, \mbox{low } \omega
\end{array}$
&
$\begin{array}{l}
2\mbox{ }\psi\mbox{'s}
\end{array}$
&
$\begin{array}{l}
\cite{Karbach_psinon,Muller,Biegel,Kohno}
\end{array}$\\
\br
\multicolumn{4}{l}
{\begin{minipage}[l]{12.3cm}{\tiny 1) One of the $\psi^*$'s contributes little to the spectral weight. Both $I_j$'s of the $\psi^*$'s are positive for $0\le k\le\pi$.
\vspace{-1.8mm}\\
2) $I_j$'s of the $\psi^*$'s have opposite signs. \quad 3) The $\psi^*$ contributes little to the spectral weight.
\vspace{-1.8mm}\\
4) This continuum has considerable spectral weight only in very low fields.}\end{minipage}}
\end{tabular}\\
\end{table}
\end{center}

\section{Summary and discussion}
We consider $S^z$ carried by the QPs: Noting that the differences of $S^z$ between the ground state and excited states of 
$S^{{\bar\alpha}\alpha}(k,\omega)$ for $\alpha=+$, $z$, and $-$ are $+1$, $0$, $-1$ 
and that 2$\psi$, $\psi\psi^*$, and 2$\psi^*$ excitations are dynamically dominant, 
respectively, we can naturally assign $S^z=+1/2$ and $-1/2$ to $\psi$ and $\psi^*$. 
Also, noting that $\sigma_2\psi\psi^*$ and $\sigma_2\psi\psi$ excitations in 2-string solutions are dominant for $S^{+-}(k,\omega)$ and $S^{zz}(k,\omega)$, 
respectively, we can naturally assign $S^z=-1$ to $\sigma_2$. 
\par
In summary, spectral properties of the spin-1/2 antiferromagnetic Heisenberg chain in a magnetic field have been investigated 
using Bethe-ansatz solutions including 2- and 3-string solutions. We identified origins of continua in terms of QPs in a magnetic field, 
$\psi$, $\psi^*$, $\sigma_2$, and $\sigma_3$. The experimentally observed high-energy peak in CPC can be understood 
as the continuum of $\sigma_2\psi\psi^*$ excitations in 2-string solutions for $S^{+-}(k,\omega)$. 
Details on contributions to sum rules and comparisons with experimental results have been shown in Ref. \cite{Kohno}. 

\ack
I am grateful to L. Balents, M. Arikawa, M. Shiroishi, and M. Takahashi for discussions, 
helpful comments and suggestions. 
This work was supported by KAKENHI 20740206 and 20046015, 
and World Premier International Research Center Initiative (WPI Initiative), MEXT, Japan.

\end{document}